\documentclass[preprint,12pt,superscriptaddress]{revtex4}  

\usepackage{graphicx} 
\usepackage{dcolumn} 
\usepackage{amsmath}  
\usepackage{amsmath}
\usepackage{amssymb}
\usepackage[percent]{overpic}
\RequirePackage{xspace}
\usepackage{relsize}
\usepackage{float}
\RequirePackage{lineno}

\graphicspath{{ps}}

\def\ra     {\!\rightarrow\!}
\def\infb     {fb$^{-1}$}
\def\evtgen   {\mbox{\textsc{EvtGen}}\xspace}
\def\geant    {\mbox{\textsc{Geant3}}\xspace}
\def\photos   {\mbox{\textsc{Photos}}\xspace}
\begin{document}
\DeclareGraphicsRule{.1}{mps}{*}{}


\preprint{\vbox{ \hbox{   }
		 \hbox{KEK Preprint 2015-64}
                 \hbox{UCHEP-16-01}
}}
\vspace*{-0.60in}

\title{ \quad\\[1.0cm] Search for the decay $\boldsymbol{B^0 \ra \phi \gamma}$}

\noaffiliation
\affiliation{University of the Basque Country UPV/EHU, 48080 Bilbao}
\affiliation{Beihang University, Beijing 100191}
\affiliation{University of Bonn, 53115 Bonn}
\affiliation{Budker Institute of Nuclear Physics SB RAS, Novosibirsk 630090}
\affiliation{Faculty of Mathematics and Physics, Charles University, 121 16 Prague}
\affiliation{Chonnam National University, Kwangju 660-701}
\affiliation{University of Cincinnati, Cincinnati, Ohio 45221}
\affiliation{Deutsches Elektronen--Synchrotron, 22607 Hamburg}
\affiliation{Gifu University, Gifu 501-1193}
\affiliation{SOKENDAI (The Graduate University for Advanced Studies), Hayama 240-0193}
\affiliation{Hanyang University, Seoul 133-791}
\affiliation{University of Hawaii, Honolulu, Hawaii 96822}
\affiliation{High Energy Accelerator Research Organization (KEK), Tsukuba 305-0801}
\affiliation{IKERBASQUE, Basque Foundation for Science, 48013 Bilbao}
\affiliation{Indian Institute of Technology Bhubaneswar, Satya Nagar 751007}
\affiliation{Indian Institute of Technology Guwahati, Assam 781039}
\affiliation{Indian Institute of Technology Madras, Chennai 600036}
\affiliation{Indiana University, Bloomington, Indiana 47408}
\affiliation{Institute of High Energy Physics, Chinese Academy of Sciences, Beijing 100049}
\affiliation{Institute of High Energy Physics, Vienna 1050}
\affiliation{J. Stefan Institute, 1000 Ljubljana}
\affiliation{Kanagawa University, Yokohama 221-8686}
\affiliation{Institut f\"ur Experimentelle Kernphysik, Karlsruher Institut f\"ur Technologie, 76131 Karlsruhe}
\affiliation{Department of Physics, Faculty of Science, King Abdulaziz University, Jeddah 21589}
\affiliation{Korea Institute of Science and Technology Information, Daejeon 305-806}
\affiliation{Korea University, Seoul 136-713}
\affiliation{Kyungpook National University, Daegu 702-701}
\affiliation{\'Ecole Polytechnique F\'ed\'erale de Lausanne (EPFL), Lausanne 1015}
\affiliation{P.N. Lebedev Physical Institute of the Russian Academy of Sciences, Moscow 119991}
\affiliation{Faculty of Mathematics and Physics, University of Ljubljana, 1000 Ljubljana}
\affiliation{Ludwig Maximilians University, 80539 Munich}
\affiliation{University of Maribor, 2000 Maribor}
\affiliation{Max-Planck-Institut f\"ur Physik, 80805 M\"unchen}
\affiliation{School of Physics, University of Melbourne, Victoria 3010}
\affiliation{Middle East Technical University, 06531 Ankara}
\affiliation{Moscow Physical Engineering Institute, Moscow 115409}
\affiliation{Moscow Institute of Physics and Technology, Moscow Region 141700}
\affiliation{Graduate School of Science, Nagoya University, Nagoya 464-8602}
\affiliation{Kobayashi-Maskawa Institute, Nagoya University, Nagoya 464-8602}
\affiliation{Nara Women's University, Nara 630-8506}
\affiliation{National United University, Miao Li 36003}
\affiliation{Department of Physics, National Taiwan University, Taipei 10617}
\affiliation{H. Niewodniczanski Institute of Nuclear Physics, Krakow 31-342}
\affiliation{Niigata University, Niigata 950-2181}
\affiliation{University of Nova Gorica, 5000 Nova Gorica}
\affiliation{Novosibirsk State University, Novosibirsk 630090}
\affiliation{Osaka City University, Osaka 558-8585}
\affiliation{Pacific Northwest National Laboratory, Richland, Washington 99352}
\affiliation{University of Pittsburgh, Pittsburgh, Pennsylvania 15260}
\affiliation{University of Science and Technology of China, Hefei 230026}
\affiliation{Seoul National University, Seoul 151-742}
\affiliation{Showa Pharmaceutical University, Tokyo 194-8543}
\affiliation{Soongsil University, Seoul 156-743}
\affiliation{University of South Carolina, Columbia, South Carolina 29208}
\affiliation{Sungkyunkwan University, Suwon 440-746}
\affiliation{School of Physics, University of Sydney, New South Wales 2006}
\affiliation{Department of Physics, Faculty of Science, University of Tabuk, Tabuk 71451}
\affiliation{Tata Institute of Fundamental Research, Mumbai 400005}
\affiliation{Excellence Cluster Universe, Technische Universit\"at M\"unchen, 85748 Garching}
\affiliation{Department of Physics, Technische Universit\"at M\"unchen, 85748 Garching}
\affiliation{Toho University, Funabashi 274-8510}
\affiliation{Department of Physics, Tohoku University, Sendai 980-8578}
\affiliation{Earthquake Research Institute, University of Tokyo, Tokyo 113-0032}
\affiliation{Department of Physics, University of Tokyo, Tokyo 113-0033}
\affiliation{Tokyo Institute of Technology, Tokyo 152-8550}
\affiliation{Tokyo Metropolitan University, Tokyo 192-0397}
\affiliation{CNP, Virginia Polytechnic Institute and State University, Blacksburg, Virginia 24061}
\affiliation{Wayne State University, Detroit, Michigan 48202}
\affiliation{Yamagata University, Yamagata 990-8560}
\affiliation{Yonsei University, Seoul 120-749}
  \author{Z.~King}\affiliation{University of Cincinnati, Cincinnati, Ohio 45221} 
  \author{B.~Pal}\affiliation{University of Cincinnati, Cincinnati, Ohio 45221} 
  \author{A.~J.~Schwartz}\affiliation{University of Cincinnati, Cincinnati, Ohio 45221} 
  \author{I.~Adachi}\affiliation{High Energy Accelerator Research Organization (KEK), Tsukuba 305-0801}\affiliation{SOKENDAI (The Graduate University for Advanced Studies), Hayama 240-0193} 
  \author{H.~Aihara}\affiliation{Department of Physics, University of Tokyo, Tokyo 113-0033} 
  \author{S.~Al~Said}\affiliation{Department of Physics, Faculty of Science, University of Tabuk, Tabuk 71451}\affiliation{Department of Physics, Faculty of Science, King Abdulaziz University, Jeddah 21589} 
  \author{D.~M.~Asner}\affiliation{Pacific Northwest National Laboratory, Richland, Washington 99352} 
  \author{H.~Atmacan}\affiliation{Middle East Technical University, 06531 Ankara} 
  \author{T.~Aushev}\affiliation{Moscow Institute of Physics and Technology, Moscow Region 141700} 
  \author{R.~Ayad}\affiliation{Department of Physics, Faculty of Science, University of Tabuk, Tabuk 71451} 
% ...
\collaboration{The Belle Collaboration}

\begin{abstract}
We have searched for the decay $B^{0} \ra \phi\gamma$ using 
the full Belle data set of $772\times 10^6$ $B\overline{B}$
pairs collected at the $\Upsilon$(4S) resonance with the
Belle detector at the KEKB $e^{+}e^{-}$ collider. No signal is
observed, and we set an upper limit on the branching fraction
of $\mathcal{B}(B^{0} \ra \phi\gamma) < 1.0 \times 10^{-7}$
at 90\% confidence level. This is the most stringent limit
on this decay mode to date.
\end{abstract}


\maketitle

{\renewcommand{\thefootnote}{\fnsymbol{footnote}}}
\setcounter{footnote}{0}

\begin{figure}[t]
\includegraphics[scale=0.62]{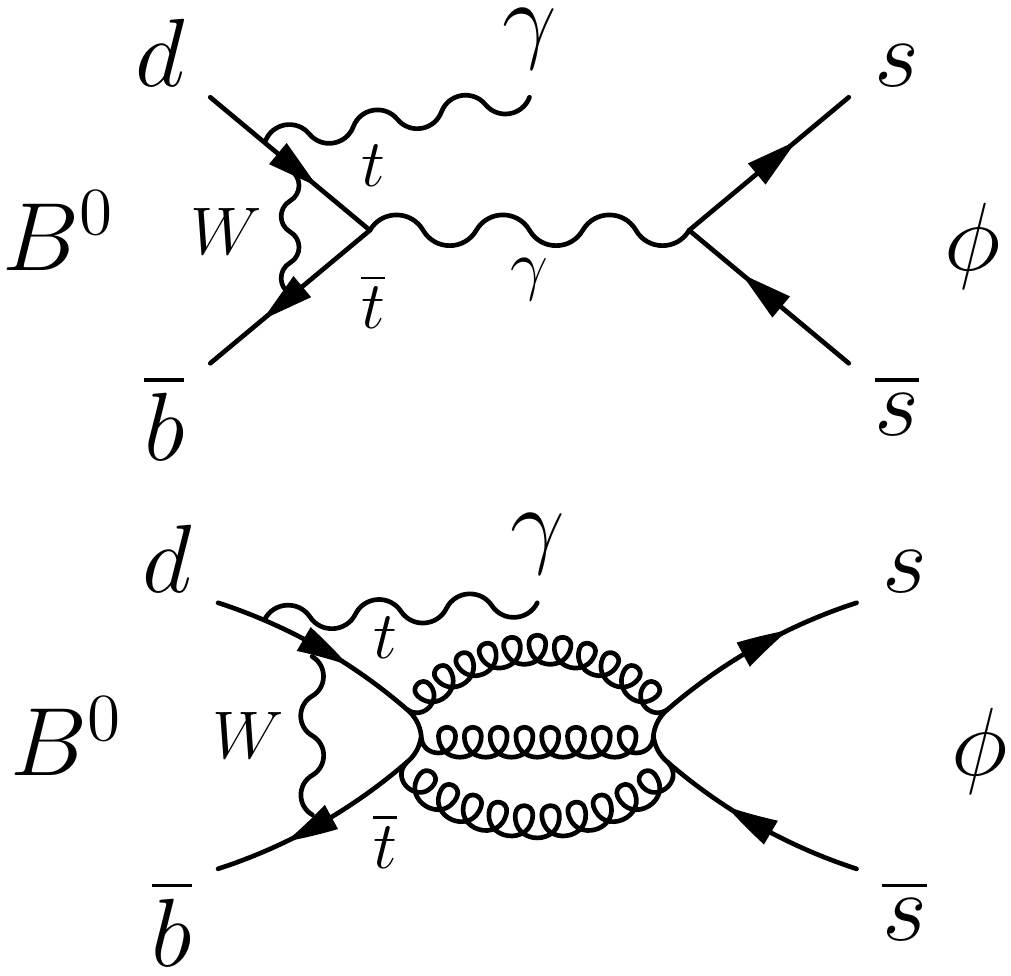}
\caption{Electroweak penguin (top) and gluonic penguin (bottom) 
contributions to $B^0 \ra \phi\gamma$.}
\label{feynman}
\end{figure}

In the Standard Model (SM), the decay
$B^{0} \ra \phi\gamma$~\cite{charge-conjugates} proceeds through
electroweak and gluonic $b\ra d$ penguin annihilation processes
as shown in Fig.~\ref{feynman}. These amplitudes are proportional
to the small Cabibbo-Kobayashi-Maskawa~\cite{CKM} matrix element
$V_{td}$ and thus are highly suppressed.
The branching fraction has been estimated based on naive QCD 
factorization~\cite{QCD} and perturbative QCD~\cite{pQCD}
and found to be in the range $10^{-12}$ to $10^{-11}$.
However, the internal loop can also be mediated by non-SM
particles such as a charged Higgs boson or supersymmetric squarks,
and thus the decay is sensitive to new physics (NP). It is estimated
that such NP could enhance the branching fraction to the level of 
$10^{-9}$ to $10^{-8}$~\cite{QCD}.
Experimentally, no evidence for this decay has been found, and the 
current upper limit on the branching fraction is $8.5 \times 10^{-7}$
at 90\% confidence level (C.L.)~\cite{babar_phig}.
Here, we present a search for this decay using the full Belle
data set of 711~\infb\ recorded on the  $\Upsilon$(4S) resonance.
This integrated luminosity corresponds to 
$(772 \pm 11) \times 10^{6}$ $B \overline{B}$ pairs, which 
is more than six times the amount of data used previously
to search for this mode.

The Belle experiment ran at the KEKB asymmetric-energy $e^+e^-$
collider located at the KEK laboratory~\cite{KEKB}. The detector
is a large-solid-angle
magnetic spectrometer consisting of a silicon vertex detector (SVD),
a 50-layer central drift chamber (CDC), an array of aerogel threshold
\u{C}erenkov counters (ACC), a barrel-like arrangement of time-of-flight
scintillation counters (TOF), and an electromagnetic calorimeter (ECL)
comprising  CsI(Tl) crystals. These detector components are 
located inside a superconducting solenoid coil that provides a 1.5~T
magnetic field.  An iron flux-return located outside the coil (KLM)
is instrumented to detect $K_L^0$ mesons and to identify muons.
Two inner detector configurations were used: a 2.0 cm beampipe
and a three-layer SVD were used for the first 140~\infb\ of data, 
while a 1.5 cm beampipe, a four-layer SVD, and a small-cell inner
drift chamber were used for the remaining 571~\infb\ of data.
The detector is described in detail elsewhere~\cite{belle_detector,SVD2}.  

Candidate photons are required to have a momentum in the range
$[2.0, 2.8]$~GeV/$c$ in the $\Upsilon(4\rm{S})$ center-of-mass (CM)
frame. To reject neutral hadrons, the photon energy deposited
in the $3\times 3$ array of ECL crystals centered on the crystal
with the highest energy must exceed 80\% of the energy deposited
in the corresponding $5\times 5$ array of crystals. 
To reduce background from $\pi^{0} \ra \gamma\gamma$ and
$\eta \ra \gamma\gamma$ decays, we pair each photon candidate
with all other photons in the event and, for each pairing,
calculate $\pi^{0}$ and $\eta$ likelihoods based on the
invariant mass. We subsequently require these likelihoods
to be less than 0.6, which preserves 97\% of the signal
while reducing the background by a factor of two.

Candidate $\phi$ mesons are reconstructed via $\phi \ra K^+K^-$
decays. Charged tracks are required to have a distance-of-closest-approach
with respect to the interaction point of less than 3.0~cm along the 
$z$ axis (anti-parallel to the $e^+$ beam), and of less than 0.3~cm in the
transverse plane. Kaons are identified using information from the
CDC, TOF, and ACC detectors. This information is used to calculate
relative likelihoods for hadron identification. A charged track with
a likelihood ratio of
${\mathcal{L}_{K}}/({\mathcal{L}_{\pi}+\mathcal{L}_{K}})>$ 0.6 is regarded
as a kaon, where $\mathcal{L}_K (\mathcal{L}_\pi)$ is the relative likelihood
of the track being a kaon (pion). The kaon identification efficiency is 85\% 
and the probability for a pion to be misidentified as a kaon is 7\%. 
Charged tracks that are consistent with
the muon hypothesis based on information from the CDC and KLM are
rejected, as are tracks consistent with the electron hypothesis
based on information from the CDC and ECL.
Oppositely charged kaon candidates are fit to a common vertex
and required to have a vertex $\chi^2$ less than 50.  The $K^+K^-$
invariant mass is required to be in the range $[1.000, 1.039]$~GeV/$c^2$,
which corresponds to $4.5\sigma$ in resolution around the $\phi$
mass~\cite{PDG}.

Candidate $B$ mesons are identified using a modified beam-energy-constrained
mass $M_{\rm bc} = \sqrt{E^2_{\rm beam} - |\vec{p}^{}_B c|^2}/c^2$, and the energy
difference $\Delta E = E_B - E_{\rm beam}$, where $E_{\rm beam}$ is the beam
energy and $\vec{p}^{}_B$ and $E_B$ are the momentum and energy, respectively,
of the $B^0$ candidate. All quantities are evaluated in the CM frame. 
To improve the $M_{\rm bc}$ resolution, the momentum $\vec{p}^{}_B$ is
calculated as
$\vec{p}^{}_{\phi} + 
(\vec{p}^{}_\gamma/|p^{}_{\gamma}|)\sqrt{(E_{\rm beam} - E_{\phi})^{2}}/c$,
where $\vec{p}^{}_{\gamma}$ is the photon momentum and $\vec{p}^{}_\phi$ 
and $E_{\phi}$ are the momentum and energy, respectively, 
of the $\phi$ candidate. We require that events satisfy 
$M_{\rm bc}\in [5.25, 5.29]~{\rm GeV}/c^2$ and 
$\Delta E\in [-0.30, 0.15]~{\rm GeV}$. The signal
yield is calculated in a smaller region  
$M_{\rm bc}\in [5.27, 5.29]~{\rm GeV}/c^2$ and 
$\Delta E\in [-0.20, 0.10]~{\rm GeV}$.

After applying the above selection criteria, less than 1\% of
events contain multiple $B$ candidates. For these events
we retain only the candidate that minimizes the difference
$|M^{}_{K^+K^-}-M^{}_\phi|$. If there remains a choice of
photons to be paired with the $\phi$, we choose the one
with the highest energy. According to Monte Carlo (MC)
simulations, these criteria select the correct $B$
candidate 96\% of the time.

Charmless hadronic decays suffer from large backgrounds arising
from continuum $e^+e^- \ra q\overline{q}\; (q = u, d, s, c)$
production. To suppress this background, we use a multivariate
analyzer based on a neural network (NN)~\cite{Feindt:2006pm}. The NN uses
the event topology and $B$-flavor-tagging information~\cite{FlavorTagging}
to discriminate continuum events, which tend to be jet-like, from
$B\overline{B}$ events, which tend to be spherical. The event shape
variables include a set of 16 modified Fox-Wolfram moments~\cite{SFW};
the cosine of the angle between the $z$ axis and the $B$ flight direction;
and the cosine of the angle between the $B$ thrust axis~\cite{NNVar} and
the thrust axis of the non-$B$-associated tracks in the event.
All of these quantities are evaluated in the CM frame. 

The NN technique requires a training procedure. 
For this training we use signal and continuum MC events. 
The MC samples are obtained using \evtgen~\cite{Lange:2001uf}
for event generation and \geant~\cite{geant3} for modeling
the detector response. Final-state radiation is taken into
account using \photos~\cite{Golonka:2005pn}.
The NN generates an output variable $C_{\rm NN}$, which ranges from
$-1$ for background-like events to $+1$ for signal-like events.
We require $C_{\rm NN} > 0.3$, which rejects 89\% of continuum
background while retaining 85\% of the signal. We then translate
$C_{\rm NN}$ to $C'_{\rm NN}$, defined as
\begin{eqnarray}
C'_{\rm NN} & = & \ln\left(\frac{C_{\rm NN} - C_{\rm min}}
{C_{\rm max} - C_{\rm NN}}\right)\,,
\end{eqnarray}
where $C_{\text{min}}$ = 0.3 and $C_{\text{max}}$ = 1.0. This translation is
convenient, as the $C'_{\rm NN}$ distribution for both signal and background
is well-modeled by a sum of Gaussian functions.

After the above selections, 961 events remain.
The remaining background consists of continuum events and rare
charmless $b$-decay processes. The latter shows peaking structure
in the $M_{\rm bc}$ distribution, with the dominant contribution
coming from $B\ra K_1(1270)\gamma,\,K_1(1270)\ra K\pi\pi$ decays.
From a large MC study we find a negligible background
contribution from $b \ra c$ processes.

We calculate signal yields using an unbinned extended maximum
likelihood fit to the observables $M_{\rm bc}$, $\Delta E$, $C'_{\rm NN}$,
and $\cos\theta_\phi$. The helicity angle $\theta_\phi$ is the angle
between the $K^+$ momentum and the opposite of the $B$ flight direction
in the $\phi$ rest frame. This variable provides additional discrimination
between signal and continuum events. The likelihood function $\mathcal{L}$
is defined as 
\begin{equation}
e^{-\sum_j Y_j} \prod_i^N \left( \sum_j Y_j 
\mathcal{P}_j(M_{\rm bc}^i, \Delta E^i, C'^i_{\rm NN}, \cos\theta_\phi^i )\right),
\end{equation} 
where $N$ is the number of candidate events (961),
$\mathcal{P}_j( M_{\rm bc}^i, \Delta E^i, C'^i_{\rm NN}, {\rm{cos}}\,\theta_\phi^i )$
is the probability density function (PDF) of component $j$ for event $i$,
and $j$ runs over all signal and background components. The parameter
$Y_j$ is the fitted yield of component~$j$.
These yields are the only free parameters in the fit.

All PDFs are obtained from MC simulation studies.
Correlations among the fit variables are found to be small, except
for a correlation between $M_{\rm bc}$ and $\Delta E$ for the charmless
background. Thus, except for this background, we factorize the PDFs as 
\begin{eqnarray}
\mathcal{P}_j(M_{\rm bc}, \Delta E, C'_{\rm NN}, {\rm cos}\,\theta_\phi) & = & \nonumber \\
 & & \hskip-1.5in
\mathcal{P}_j(M_{\rm bc})\cdot\mathcal{P}_j(\Delta E)\cdot\mathcal{P}_j(C'_{\rm NN})\cdot
\mathcal{P}_j(\cos\theta_\phi).
\end{eqnarray}
The $M_{\rm bc}$ and $\Delta E$ distributions for signal are
modeled with Crystal Ball functions~\cite{crystalball}, while
the $C'_{\rm NN}$ and $\cos\theta_\phi$ distributions are
modeled with a bifurcated Gaussian and the function
$1 - \cos^2\theta_\phi$, respectively. The peak positions
and resolutions of the $M_{\rm bc}$, $\Delta E$, and $C'_{\rm NN}$
PDFs are adjusted to account for small data-MC differences
observed in a high-statistics control sample of 
$B^0 \ra K^{*0}(\ra K^+\pi^-)\gamma$ decays, which
have a similar topology as $B^0 \ra \phi \gamma$. 

For the charmless 
background, the $C'_{\rm NN}$ component
is modeled with a Gaussian function. The
peak position and resolution are adjusted from
data-MC differences observed for the 
charmless background
in the $B^0 \ra K^{*0}(\ra K^+\pi^-)\gamma$ control sample.
The $M_{\rm bc}$ and $\Delta E$ components are modeled 
by a joint two-dimensional non-parametric function based on
kernel estimation~\cite{Cranmer:2000du}, to account for
their correlation. The $\cos\theta_\phi$ distribution is
modeled by a one-dimensional non-parametric function.
For continuum background, the $M_{\rm bc}$ shape is modeled
by an ARGUS function~\cite{Albrecht:1990am}, and the
$C'_{\rm NN}$ shape is modeled by the sum of two Gaussians
having a common mean. The peak positions and resolutions
are adjusted from data-MC differences observed for the
continuum background of the control sample. The $\Delta E$
and $\cos\theta_\phi$ distributions are modeled by
Chebyshev polynomials of the first and second order,
respectively. All shape parameters of these PDFs are
fixed to the corresponding MC values. To test the stability of the
fitting procedure, we perform numerous fits on large ensembles
of MC events; in all cases the input value is recovered within
the statistical error. 

\begin{figure*}[h!t!p!]
\hspace*{\fill}
\hbox{
\includegraphics[scale=0.34]{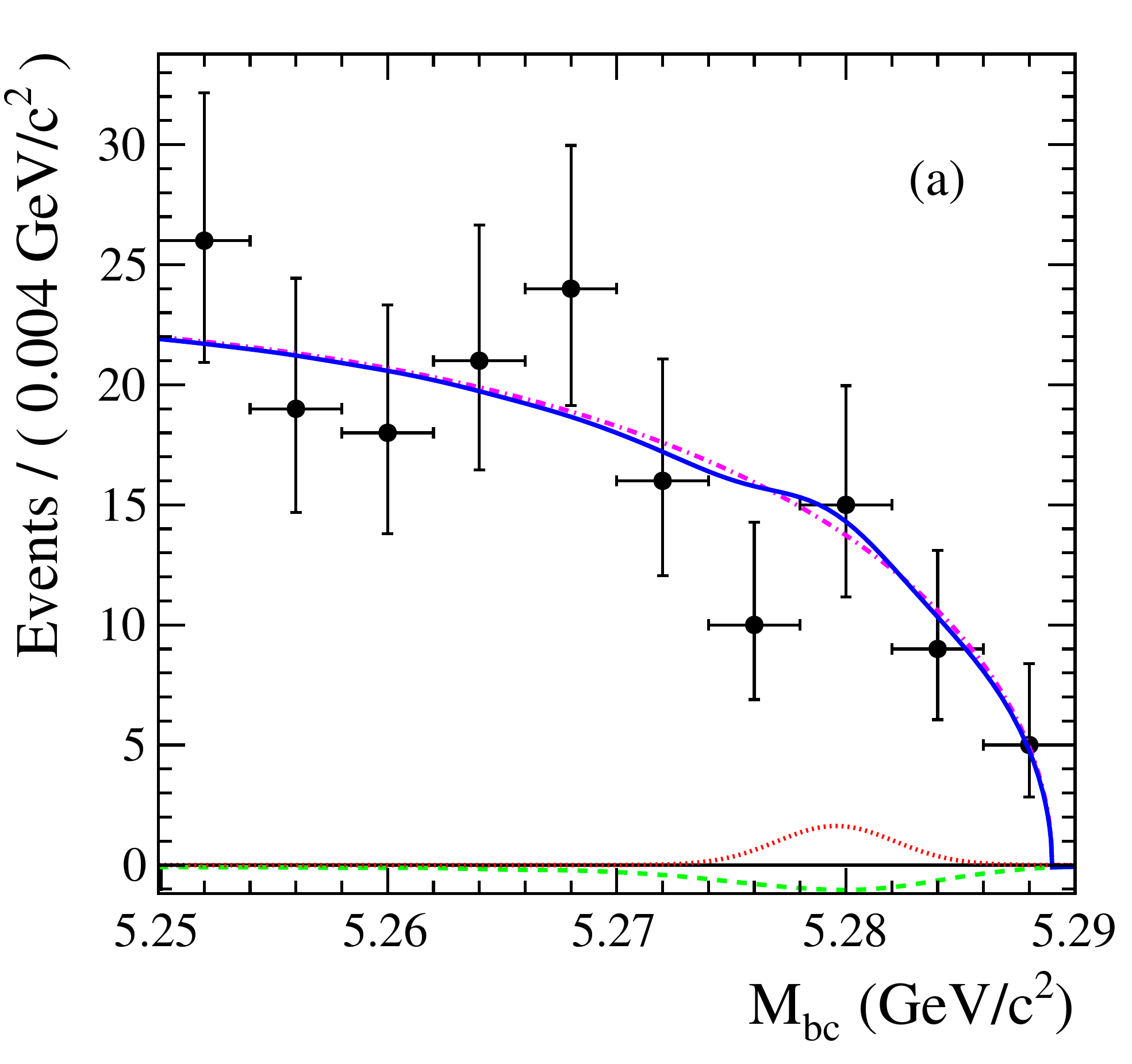}
\includegraphics[scale=0.34]{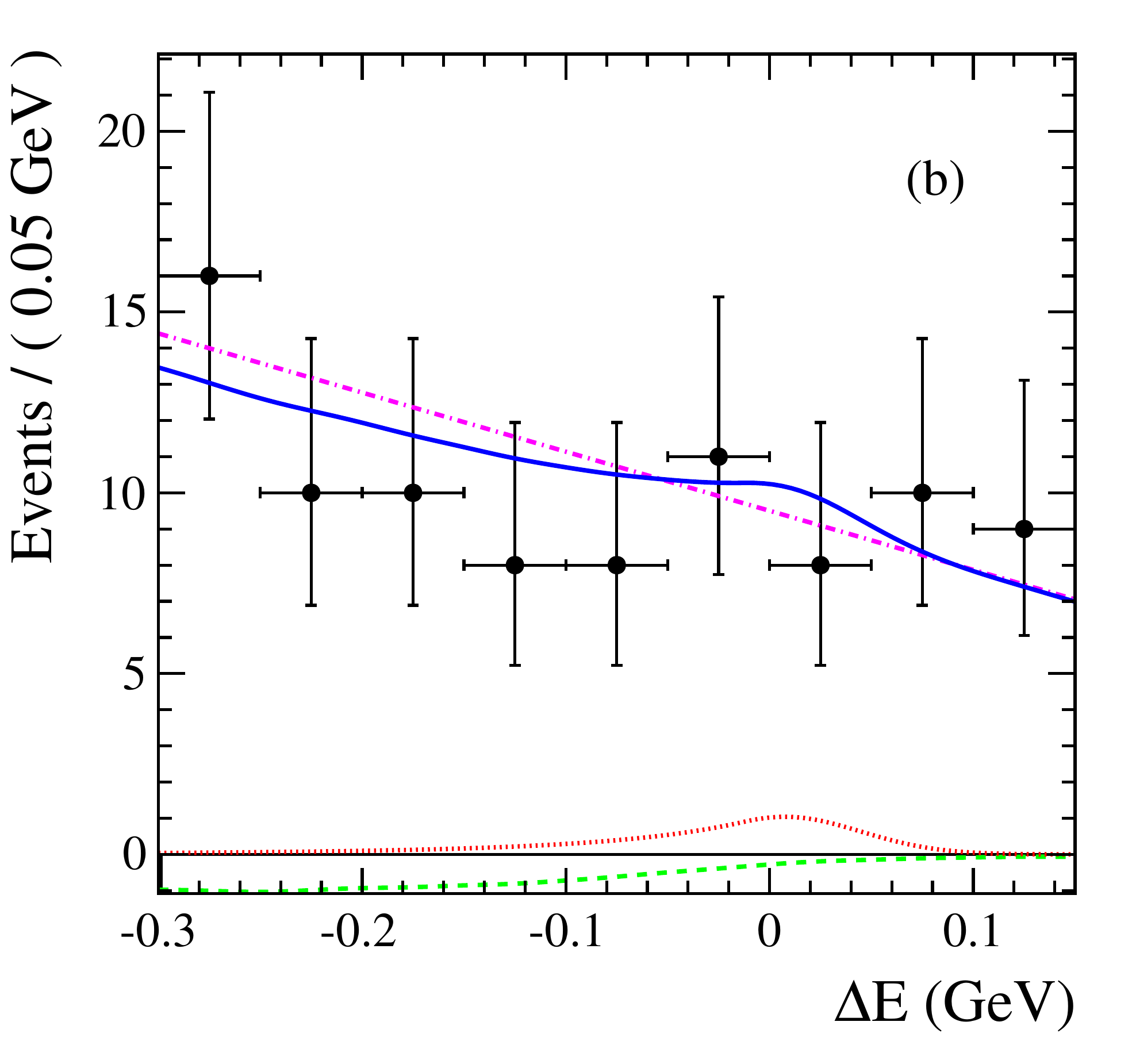}
\hskip0.50in
} 
\hbox{
\includegraphics[scale=0.34]{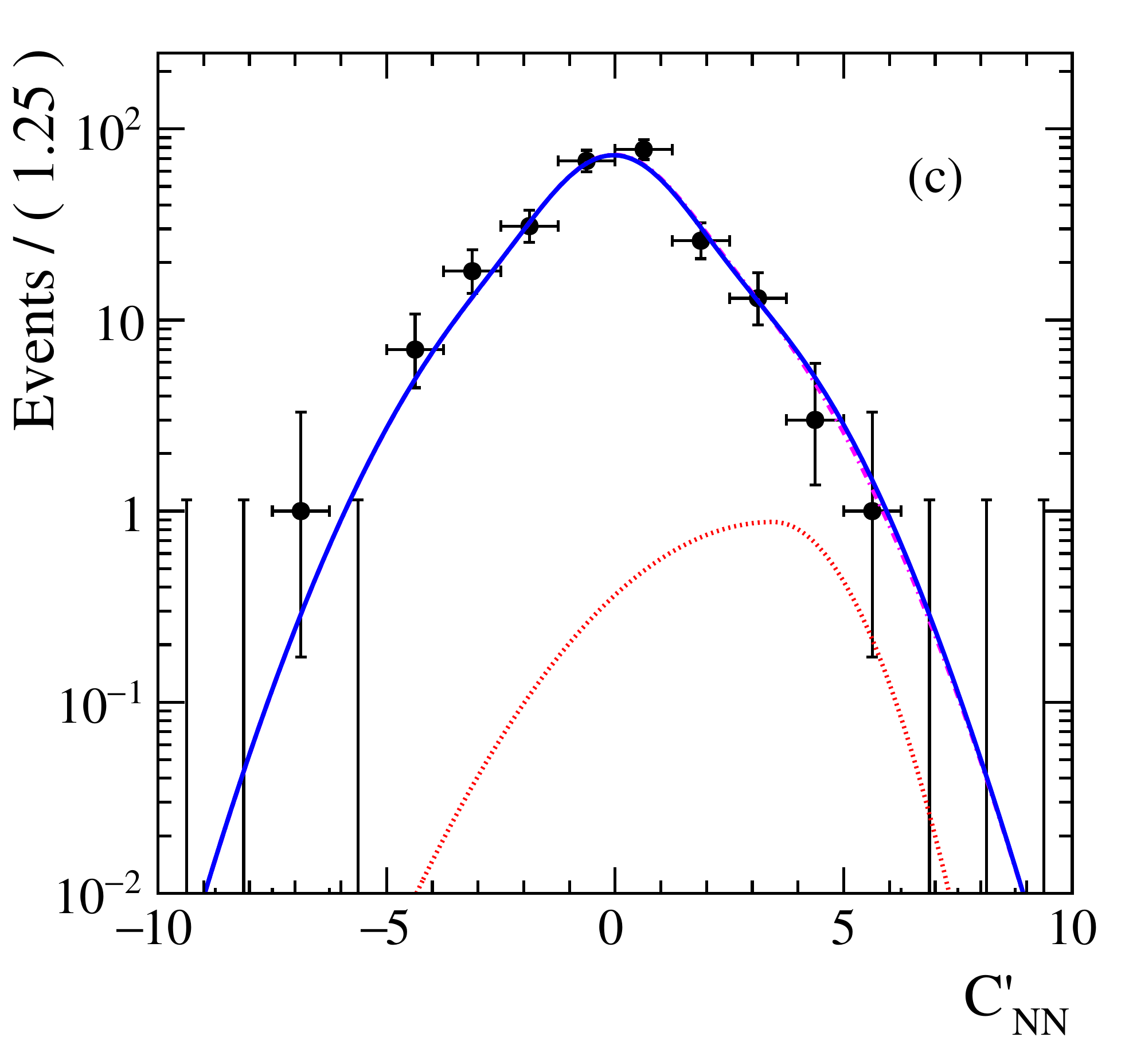}
\includegraphics[scale=0.34]{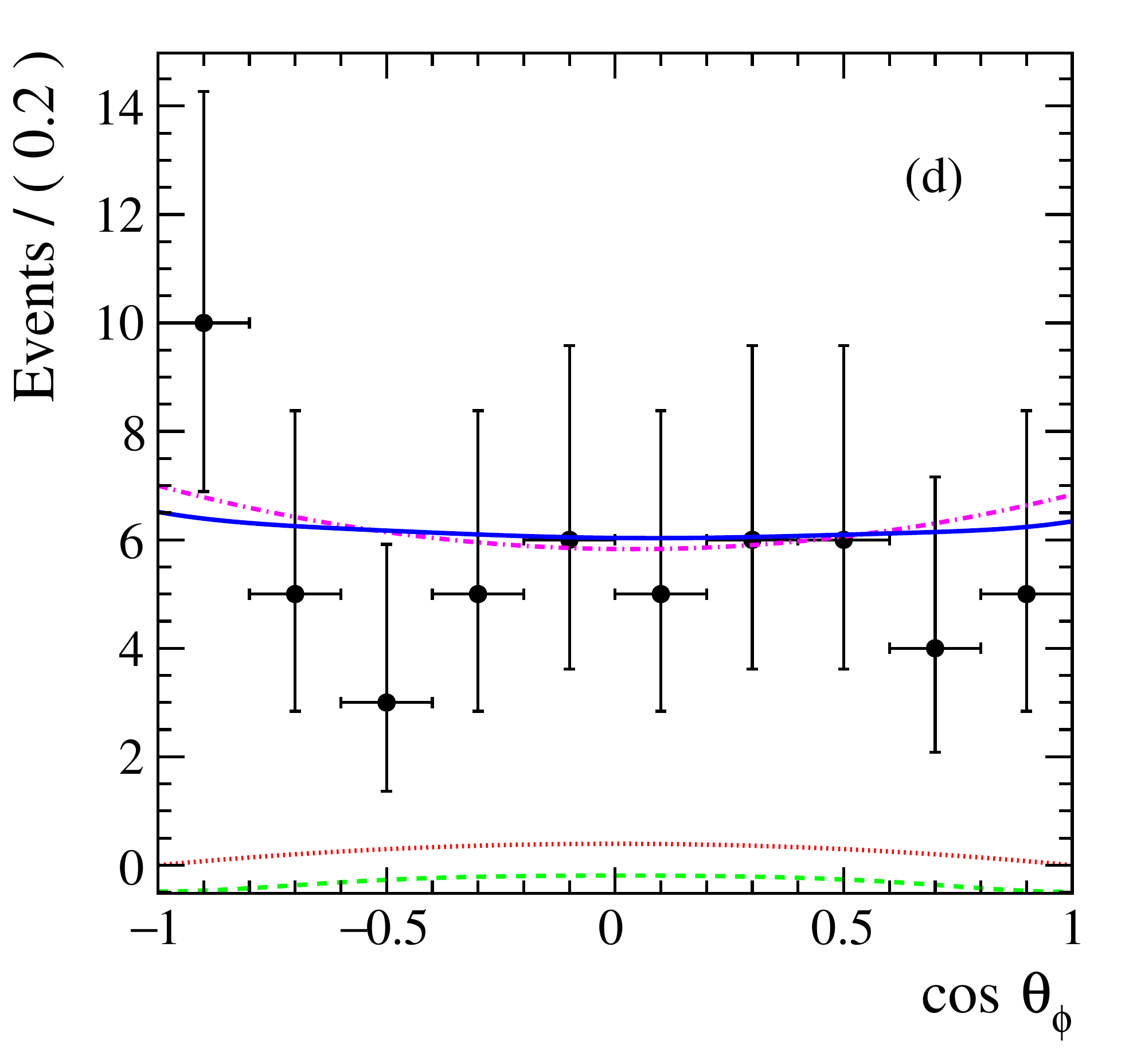}
}
\caption{Projections of the four-dimensional fit:
(a) $M_{\rm bc}$ in the $\Delta E$ signal region; 
(b) $\Delta E$ in the $M_{\rm bc}$ signal region; 
(c) $C'_{\rm NN}$ in the $M_{\rm bc}$ and $\Delta E$ signal regions; and 
(d) ${\rm cos}\,\theta_\phi$ in the $M_{\rm bc}$ and $\Delta E$ signal
regions. 
Plots (a), (b), and (d) also require $C'_{\rm NN} > 1$.
The points with error bars show the data;
the dotted (red) curves represent the signal;
the dashed-dotted (magenta) curves represent continuum events; 
the dashed (green) curves represent the charmless background;
and the solid (blue) curves represent the total.}
\label{datafit}
\end{figure*}

The projections of the fit are shown in Fig.~\ref{datafit}. 
The resulting branching fraction is calculated as
\begin{eqnarray}
\mathcal{B}\left(B^0\ra \phi \gamma \right) & = & 
\frac{Y_{\rm sig}}
{N_{B\overline{B}} \cdot \varepsilon \cdot \mathcal{B}(\phi \ra K^+K^-)}\,,
\label{eqn:br}
\end{eqnarray}
where $Y_{\rm sig} = 3.4\,^{+4.6}_{-3.8}$ is the 
signal yield in the signal region;
$\varepsilon = 0.296\pm 0.001$ is the signal efficiency 
in this region as calculated from MC simulation;
$N_{B\overline{B}} = (772 \pm 11) \times 10^6$ is the number
of $B\overline{B}$ events; and 
$\mathcal{B}(\phi \ra K^+K^-) = (48.9 \pm 0.5)\%$ is
the branching fraction for $\phi\ra K^+K^-$~\cite{PDG}.
The efficiency $\varepsilon$ is corrected by a factor $1.024 \pm 0.010$
to account for a small difference in particle identification efficiencies
between data and simulations. This correction is estimated from a sample
of $D^{*+} \ra D^0(\ra K^-\pi^+)\pi^+$ decays~\cite{pid_belle}.
In Eq.~(\ref{eqn:br}) we assume equal production of
$B^0\overline{B}{}^{\,0}$ and $B^+B^-$ pairs at the
$\Upsilon$(4S) resonance.

We observe no statistically significant signal and  
set an upper limit on the number of signal events
by integrating the area under the likelihood function
${\cal L}(Y^{}_{\rm sig})$.
The value of $Y^{}_{\rm sig}$
that corresponds to 90\% of the total area from zero to
infinity is taken as the 90\% C.L. upper limit~\cite{bayesian_limit}.
This value is converted to an upper limit on the branching
fraction ${\cal B}$ using Eq.~(\ref{eqn:br}); the result is
\begin{eqnarray}
{\cal B}(B^0 \ra \phi \gamma) & < & 1.0 \times 10^{-7}\,.
\end{eqnarray}
We include systematic uncertainties (discussed below) in the upper
limit by convolving the likelihood function with a Gaussian function
whose width is set equal to the total systematic uncertainty. 
We perform this convolution before calculating the upper
limit on $Y^{}_{\rm sig}$.

The systematic uncertainties on the branching fraction 
are listed in Table~\ref{tab:sys}. 
The largest uncertainty is due to the fixed parameters in the PDFs.
We evaluate this by varying each parameter individually according
to its statistical uncertainty. The resulting changes in
$Y^{}_{\rm sig}$ are added in quadrature to obtain the systematic
uncertainty. We evaluate, in a similar manner, the uncertainty due
to errors in the calibration factors. The sum in quadrature of
these two uncertainties is listed in Table~\ref{tab:sys} as the
uncertainty due to PDF parameterization.

To test for potential bias in our fitting procedure,
we fit a large ensemble of MC events. By comparing the mean of the 
yields obtained with the input value, a potential bias of $-0.08$ event
is found. We attribute this to neglecting small correlations between
the fitted variables and take this bias as a systematic uncertainty.
The uncertainty due to the $C_{\rm NN}$ selection is determined
by applying different $C_{\rm NN}$ criteria to the control sample;
the difference in the changes observed between data and MC simulation
is taken as the systematic uncertainty. The uncertainty due to the background
sample used in training the NN is determined by changing the training
sample and noting the change in the signal yield of the control sample. 
The systematic uncertainty due to charged track reconstruction
is determined from a study of partially reconstructed
$D^{*+}\ra D^0 (\ra K^0_S\pi^+\pi^-)\pi^+$ decays
and found to be 0.35\% per track. 
An uncertainty due to particle identification of 
0.8\% per kaon is obtained from a study of $D^{*+}\ra D^0 (\ra K^-\pi^+)\pi^+$
decays. The uncertainty on $\varepsilon$ due to MC statistics is 0.2\%,
and the uncertainty on the number of $B \overline {B}$ pairs is 1.4\%.
The total systematic uncertainty is obtained by summing all individual
contributions in quadrature; the result 
corresponds to $\pm 1.2$ events.

\begin{table}[t]
\renewcommand{\arraystretch}{1.0} 
\caption{\small Systematic uncertainties on $\mathcal{B}(B^0 \ra \phi \gamma)$
in units of number of events. We convert fractional errors to number of events
for easy comparison. Uncertainties listed in the lower section are external to
our analysis. }
\label{tab:sys}
\begin{tabular}{ c | c } 
\hline \hline
Source & Uncertainty (events) \\ \hline
PDF parameterization & $^{+1.21}_{-1.14}$ \\ 
Fit bias & $^{+0.00}_{-0.08}$ \\
$C_{\rm NN}$ selection efficiency & 0.03 \\
$C_{\rm NN}$ background sample & 0.02 \\
Tracking efficiency & 0.02 \\
PID efficiency & 0.05\\
Photon reconstruction & 0.08 \\
MC statistics & 0.01 \\
\hline
$\mathcal{B}(\phi \ra K^{+}K^{-})$ & 0.03\\
Number of $B\overline{B}$ events & 0.05 \\ 
\hline
Total & $^{+1.22}_{-1.15}$ \\ 
\hline \hline 
\end{tabular}
\end{table}

In summary, we have searched for the decay $B^0 \ra \phi\gamma$ using the full
Belle data set. We find no evidence for this decay and set an upper limit
on the branching fraction of 
$\mathcal{B}(B^0 \ra \phi \gamma) < 1.0 \times 10^{-7}$ at 90\% C.L.
This limit is almost an order of magnitude lower than the
previous most stringent result~\cite{babar_phig}.

We thank the KEKB group for the excellent operation of the
accelerator; the KEK cryogenics group for the efficient
operation of the solenoid; and the KEK computer group,
the National Institute of Informatics, and the 
PNNL/EMSL computing group for valuable computing
and SINET4 network support.  We acknowledge support from
the Ministry of Education, Culture, Sports, Science, and
Technology (MEXT) of Japan, the Japan Society for the 
Promotion of Science (JSPS), and the Tau-Lepton Physics 
Research Center of Nagoya University; 
the Australian Research Council;
Austrian Science Fund under Grant No.~P 22742-N16 and P 26794-N20;
the National Natural Science Foundation of China under Contracts 
No.~10575109, No.~10775142, No.~10875115, No.~11175187, No.~11475187
and No.~11575017;
the Chinese Academy of Science Center for Excellence in Particle Physics; 
the Ministry of Education, Youth and Sports of the Czech
Republic under Contract No.~LG14034;
the Carl Zeiss Foundation, the Deutsche Forschungsgemeinschaft, the
Excellence Cluster Universe, and the VolkswagenStiftung;
the Department of Science and Technology of India; 
the Istituto Nazionale di Fisica Nucleare of Italy; 
the WCU program of the Ministry of Education, National Research Foundation (NRF) 
of Korea Grants No.~2011-0029457,  No.~2012-0008143,  
No.~2012R1A1A2008330, No.~2013R1A1A3007772, No.~2014R1A2A2A01005286, 
No.~2014R1A2A2A01002734, No.~2015R1A2A2A01003280 , No. 2015H1A2A1033649;
the Basic Research Lab program under NRF Grant No.~KRF-2011-0020333,
Center for Korean J-PARC Users, No.~NRF-2013K1A3A7A06056592; 
the Brain Korea 21-Plus program and Radiation Science Research Institute;
the Polish Ministry of Science and Higher Education and 
the National Science Center;
the Ministry of Education and Science of the Russian Federation and
the Russian Foundation for Basic Research;
the Slovenian Research Agency;
Ikerbasque, Basque Foundation for Science and
the Euskal Herriko Unibertsitatea (UPV/EHU) under program UFI 11/55 (Spain);
the Swiss National Science Foundation; 
the Ministry of Education and the Ministry of Science and Technology of Taiwan;
and the U.S.\ Department of Energy and the National Science Foundation.
This work is supported by a Grant-in-Aid from MEXT for 
Science Research in a Priority Area (``New Development of 
Flavor Physics'') and from JSPS for Creative Scientific 
Research (``Evolution of Tau-lepton Physics'').


\begin{thebibliography}{99}
\bibitem{charge-conjugates}
Throughout this paper, charge-conjugate decay modes are
implicitly included unless stated otherwise.
\bibitem{bayesian_limit}
As our upper limit is obtained using a Bayesian method,
it corresponds to 90\% ``credibility level.'' However,
we use the (frequentist) term ``confidence level''
following common convention.
\end{thebibliography}
\end{document}